\documentclass[12pt]{article}

\begin{document}

\begin{center}
{\Large\bf General relativistic tidal heating for M$\o$ller
pseudotensor}
\end{center}

\begin{center}
Lau Loi So\footnote{s0242010@gmail.com}
\end{center}

\begin{abstract}
Thorne elucidated that the relativistic tidal heating is the same as the
Newtonian theory.  Moreover, Thorne also claimed that the tidal
heating is independent of how one localizes gravitational energy
and is unambiguously given by a certain formula. Purdue and Favata calculated the tidal heating for different classical pseudotensors
including M{\o}ller and obtained the results all matched with the
Newtonian perspective. After re-examined this M$\o$ller
pseudotensor, we find that there does not exist any tidal heating
value. Thus we claim that the relativistic tidal heating
is pseudotensor independent under the condition that if the peusdotensor is a Freud typed superpotential.
\end{abstract}

\section{Introduction}
Tidal heating is a kind of physical phenomenon which means the net
work done by an external tidal field on an isolated body. Our
daily life experience is the tide on Earth. However, a more
dramatical example is the Jupiter-Io system~\cite{Peale}.    
In 1998 Thorne~\cite{Thorne} elucidated that the expected tidal heating
rate is the same both in relativistic and Newtonian gravity:
$\dot{W}=-\frac{1}{2}\dot{I}_{ij}E^{ij}$, where $W$ refers to the
tidal work, the dot means differentiate w.r.t. time t, $I_{ij}$ is
the mass quadrupole moment of the isolated body and $E_{ij}$ is
the tidal field of the external universe. Note that both $I_{ij}$
and $E_{ij}$ are time dependent, symmetric and trace free. 
Moreover, Thorne also claimed
that this tidal heating is independent of how one localizes the 
gravitational energy and is unambiguously gives by a certain value.  
This was being proved by calculating $\dot{W}$ explicitly using various gravitational pseudotensors.

In 1999, Purdue~\cite{Purdue} used the Landau-Lifshitz
pseudotensor~\cite{LL} to calculate the tidal heating and the
result matched with the Newtonian perspective. Later in 2001,
Favata~\cite{Favata} employed the same method to verify the
Einstein, Bergmann-Thomson and
M{\o}ller pseudotensors~\cite{Freud,BT,Moller}, who claimed that all of
them give the same result as Purdue did.   It seems have accomplished the verification that the tidal heating is pseudotensor independent.  
However, after re-examined the M$\o$ller
pseudotensor, we find zero gravitational energy and vanishing
tidal heating. It is known that the M$\o$ller pseudotensor is not a physical
acceptable candidate for calculating the gravitational
energy-momentum because it does not satisfy the Einstein field
equation inside matter~\cite{CQGSoNesterChen2009}. Here we claim
that obtaining the Einstein equation through the Freud typed superpotential 
guarantees the expected tidal heating~\cite{SoarXiv} but the
converse is not true (see section 2.3). The present paper is trying to illustrate that the relativistic tidal heating can be pseudotensor independent under the condition that the pseudotensor is a Freud typed superpotential (i.e., see (\ref{30aSep2015})).

\section{Technical background}
Here we use $\eta_{\mu\nu}=(-1,1,1,1)$ as our spacetime
signature~\cite{MTW} and we set the geometrical units $G=c=1$,
where $G$ and $c$ are the Newtonian constant and speed of light.
The Greek letters denote the spacetime and Latin letters refer to
spatial. In principle, the classical
pseudotensor~\cite{CQGSoNesterChen2009} can be obtained from a
rearrangement of the Einstein equation:
$G_{\mu\nu}=\kappa{}T_{\mu\nu}$, where constant
$\kappa=8\pi{}G/c^{4}$ and $T_{\mu\nu}$ is the stress tensor. We
define the gravitational energy-momentum pseudotensor in terms of
a suitable superpotential $U_{\alpha}{}^{[\mu\nu]}$:
\begin{eqnarray}
2\kappa\sqrt{-g}\,t_{\alpha}{}^{\mu}:=\partial_{\nu}U_{\alpha}{}^{[\mu\nu]}
-2\sqrt{-g}\,G_{\alpha}{}^{\mu}.
\label{6aMar2015}
\end{eqnarray}
The total energy-momentum density complex can be defined as
\begin{eqnarray}
{\cal{T}}_{\alpha}{}^{\mu}:=\sqrt{-g}(T_{\alpha}{}^{\mu}+t_{\alpha}{}^{\mu})
=(2\kappa)^{-1}\partial_{\nu}U_{\alpha}{}^{[\mu\nu]}.
\end{eqnarray}
Note that both inside matter and in empty spacetime satisfy the
energy-momentum conservation law:
$\partial_{\mu}T_{\alpha}{}^{\mu}=0=\partial_{\mu}t_{\alpha}{}^{\mu}$.
For the condition of interior mass-energy density, it is known
that all classical pseudotensors give the standard result
$2G^{\mu}{}_{\alpha}$, but only M$\o$ller
failed~\cite{CQGSoNesterChen2009}, i.e.,
$\partial_{\nu}U_{\alpha}{}^{\mu\nu}=R^{\mu}{}_{\alpha}$ in
Riemann normal coordinates. This M$\o$ller pseudotensor is
disqualified as a satisfactory description of energy-momentum,
thus investigate the tidal heating becomes not important.

There are infinite numbers of pseudotensors, perhaps the Freud
superpotential~\cite{Freud}
$_{F}U_{\alpha}{}^{[\mu\nu]}:=-\sqrt{-g}g^{\beta\sigma}
\Gamma^{\tau}{}_{\beta\lambda}\delta^{\lambda\mu\nu}_{\tau\sigma\alpha}$
is a simpler expression to demonstrate the tidal work. Here we
emphasize that pseudotensor not only require to satisfy the
condition of the mass-energy density at interior~\cite{MTW}, but
also the ADM mass at spatial infinity~\cite{ADM}. We know only the
Freud typed superpotential provides these two requirements
simultaneously~\cite{CQGSo2009}. We simply demonstrate this
argument by an arbitrary superpotential
\begin{eqnarray}
U_{\alpha}{}^{[\mu\nu]}:=\sqrt{-g}(
k_{1}\delta^{\rho}_{\alpha}\Gamma^{\lambda\tau}{}_{\lambda}
+k_{2}\delta^{\tau}_{\alpha}\Gamma^{\rho\lambda}{}_{\lambda}
+k_{3}\Gamma^{\tau\rho}{}_{\alpha})\delta^{\mu\nu}_{\rho\tau}.\label{30aSep2015}
\end{eqnarray}
In order to obtain the expected result inside matter and at spatial
infinity, the only possibility for constants $k_{1}$, $k_{2}$ and
$k_{3}$ are unity.

In vacuum, the quantity $t_{0}{}^{0}$ can be interpreted as the
gravitational energy and $t_{0}{}^{j}$ is the energy flux. The
tidal work can be computed as
\begin{eqnarray}
2\kappa\dot{W}=-\int_{V}\sqrt{-g}\partial_{t}t_{0}{}^{0}d^{3}x=\int_{V}\sqrt{-g}\partial_{j}t_{0}{}^{j}d^{3}x,
\label{21aSep2015}
\end{eqnarray}
where we have used the energy-momentum conservation law
$\partial_{\mu}t_{0}{}^{\mu}=0$ at the last step. Moreover, using
the Gauss's theorem, the last integral in (\ref{21aSep2015}) can
be written as a surface integral
$\oint_{\partial{}V}\sqrt{-g}\,t_{0}{}^{j}\,\hat{n}_{j}\,r^{2}\,d\Omega$.
Note that $r\equiv\sqrt{\delta_{ab}x^{a}x^{b}}$ is the distance
from the body in its local asymptotic rest frame and
$\hat{n}_{j}\equiv{}x_{j}/r$ is the unit radial vector. We adapt
the deDonder (harmonic) gauge for the tidal heating calculation:
$0=\partial_{\beta}(\sqrt{-g}g^{\alpha\beta})=\sqrt{-g}\Gamma^{\alpha\beta}{}_{\beta}$.
The metric tensor can be decomposed as
\begin{eqnarray}
g_{\mu\nu}=\eta_{\mu\nu}+h_{\mu\nu}+...,\quad{}g^{\mu\nu}=\eta^{\mu\nu}-h^{\mu\nu}+....
\end{eqnarray}
For our tidal heating calculation purpose, we only consider the
accuracy~\cite{Purdue}
\begin{eqnarray}
\dot{W}=a_{1}\partial_{t}(I_{ij}E^{ij})+a_{2}\dot{I}_{ij}E^{ij},\label{23cSep2015}
\end{eqnarray}
where $a_{1}$, $a_{2}$ are constants. Here we remark that the
coefficient $a_{1}$ indicates a specify choice for the energy
localization and $\partial_{t}(I_{ij}E^{ij})$ is a reversible
ambiguous tidal-quadrupole interaction process. Meanwhile it is
expected $a_{2}=-\frac{1}{2}$ and $-\frac{1}{2}\dot{I}_{ij}E^{ij}$
is the unambiguous irreversible tidal heating dissipation process
which we are most interested in. The related physical
expressions~\cite{Purdue} are:
\begin{eqnarray}
&&h_{00}=\frac{2M}{r}+\frac{3}{r^{5}}I_{ij}x^{i}x^{j}-E_{ij}x^{i}x^{j},~
h_{0j}=-\frac{2}{r^{3}}\dot{I}_{ij}x^{i}-\frac{10}{21}\dot{E}_{ik}x^{i}x^{k}x_{j}+\frac{4}{21}\dot{E}_{ij}x^{i}r^{2},
\quad\label{27cMar2015}
\end{eqnarray}
and $h_{ij}=\delta_{ij}h_{00}$. The value of the weighting
factor~\cite{Favata} $\sqrt{-g}=1+h_{00}+...$.

\section{Vanishing tidal heating for M$\o$ller pseudotensor}
It is straightforward to double check the tidal heating for
Einstein and M$\o$ller pseudotensors~\cite{Favata}, we examine the
difference between these two pseudotensors and we named it S. We
demonstrate that Einstein pseudotensor satisfies the expected
inside matter requirement, but both M$\o$ller and S pseudotensors
do not.

\subsection{Einstein pseudotensor}
Here we recall the result of
tidal heating for Einstein pseudotensor~\cite{Freud}. Using Freud
superpotential $_{F}U_{\alpha}{}^{[\mu\nu]}$and we have the
Einstein pseudotensor
$\partial_{\nu}({}_{F}U_{\alpha}{}^{[\mu\nu]})$.  Inside matter
gives the desired value
$2G^{\mu}{}_{\alpha}$~\cite{CQGSoNesterChen2009}. In vacuum
\begin{eqnarray}
_{E}t_{\alpha}{}^{\mu}=\delta^{\mu}_{\alpha}(\Gamma^{\beta\lambda}{}_{\nu}\Gamma^{\nu}{}_{\beta\lambda}
-\Gamma^{\pi}{}_{\pi\nu}\Gamma^{\nu\lambda}{}_{\lambda})
+\Gamma^{\nu\beta}{}_{\nu}(\Gamma^{\mu\beta}{}_{\alpha}+\Gamma^{\beta\mu}{}_{\alpha})
+\Gamma^{\pi}{}_{\pi\alpha}(\Gamma^{\mu\lambda}{}_{\lambda}-\Gamma^{\lambda\mu}{}_{\lambda})
-2\Gamma^{\beta\nu}{}_{\alpha}\Gamma^{\mu}{}_{\beta\nu}.
\end{eqnarray}
The corresponding gravitational energy density and energy flux are
\begin{eqnarray}
_{E}t_{0}{}^{0}=-\frac{1}{2}\eta^{cd}h_{00,c}h_{00,d},\quad{}
_{E}t_{0}{}^{j}=\eta^{ja}h_{00,0}h_{00,a}.
\end{eqnarray}
The known tidal heating rate is
$\dot{W}=\frac{3}{10}\partial_{t}(I_{ij}E^{ij})-\frac{1}{2}\dot{I}_{ij}E^{ij}$.

\subsection{M$\o$ller pseudotensor} Here we define the M$\o$ller
superpotential as
$_{M}U_{\alpha}{}^{[\mu\nu]}:=\sqrt{-g}(\Gamma^{\nu\mu}{}_{\alpha}
-\Gamma^{\mu\nu}{}_{\alpha})$~\cite{Moller} and the
associated pseudotensor is
\begin{eqnarray}
&&_{M}t_{\alpha}{}^{\mu}
=\Gamma^{\beta\nu}{}_{\alpha}(\Gamma_{\beta\nu}{}^{\mu}-\Gamma_{\nu\beta}{}^{\mu})
+(\Gamma^{\mu\beta}{}_{\alpha}-\Gamma^{\beta\mu}{}_{\alpha})\Gamma_{\beta}{}^{\nu}{}_{\nu}
+g^{\mu\rho}g^{\beta\nu}(g_{\alpha\beta,\rho\nu}-g_{\alpha\rho,\beta\nu})\label{17aSep2015}\\
&&=2R^{\mu}{}_{\alpha}
-(\Gamma^{\beta\nu}{}_{\alpha}+\Gamma^{\nu\beta}{}_{\alpha})\Gamma_{\beta\nu}{}^{\mu}
+(\Gamma^{\beta\mu}{}_{\alpha}+\Gamma^{\mu\beta}{}_{\alpha})\Gamma_{\beta}{}^{\nu}{}_{\nu}
+g^{\mu\rho}g^{\beta\nu}(g_{\beta\nu,\alpha\rho}-g_{\rho\beta,\alpha\nu}).\quad\quad\label{7aSep2015}
\end{eqnarray}
The value for inside matter in Riemann normal coordinates is
$R^{\mu}{}_{\alpha}$~\cite{CQGSoNesterChen2009}. In vacuum, using
(\ref{7aSep2015}) and apply the deDonder gauge condition, the
energy density is
\begin{eqnarray}
_{M}t_{0}{}^{0}
=2h_{00,0}h_{00,0}+\eta^{cd}(2h_{0c}h_{00,0d}-h_{0c,0}h_{0d,0}+g_{0c,0d}-g_{cd,00}).\label{10aSep2015}
\end{eqnarray}
Generally we expect
$t_{0}{}^{0}\sim\eta^{cd}h_{00,c}h_{00,d}$~\cite{Purdue} which
means it should not involved any time derivative, then we can
immediately justify that both the gravitational energy and tidal
heating vanish because the energy-momentum conservation
$0=\partial_{0}t_{0}{}^{0}+\partial_{j}t_{0}{}^{j}$. For the
completeness, we check the energy flux from (\ref{17aSep2015})
\begin{eqnarray}
_{M}t_{0}{}^{j}&=&\eta^{ja}(h_{00,0}h_{00,a}-4h_{00}h_{00,0a}-g_{00,0a})\nonumber\\
&&+\eta^{ja}\eta^{cd}(h_{0c}h_{00,ad}-h_{00,c}h_{0d,a}+g_{0c,ad}-g_{0a,cd}).\label{17bSep2015}
\end{eqnarray}
The vanishing tidal heating can be manipulated as
\begin{eqnarray}
0=\int_{V}\sqrt{-g}\eta^{ja}\partial_{j}\left[(h_{00,0}h_{00,a}-4h_{00}h_{00,0a}-g_{00,0a})
+\eta^{cd}(h_{0c}h_{00,ad}-h_{00,c}h_{0d,a})\right]d^{3}x.\label{23aSep2015}
\end{eqnarray}
Then we deduce the following identity using (\ref{23aSep2015})
\begin{eqnarray}
\frac{1}{2\kappa}\oint_{\partial{}V}\sqrt{-g}\eta^{ja}g_{00,0a}\hat{n}_{j}r^{2}d\Omega
=\frac{1}{10}\frac{d}{dt}(I_{ij}E^{ij})-\frac{1}{2}\dot{I}_{ij}E^{ij}.
\end{eqnarray}
For the M$\o$ller pseudotensor, Favata obtained the desired tidal
heating value but we get null.  However, we do have the same
result which is the vanishing energy density $_{M}t_{0}{}^{0}$. We
observed that equation (74) in \cite{Favata} contains time
derivative:
\begin{eqnarray}
-16\pi{}_{M}\tau_{0}{}^{0}=\sqrt{-g}(-h_{,00}-h_{00}h_{,00}-h_{,0\beta}h^{0\beta}-2h^{\alpha\sigma}h_{0\alpha,0\sigma}
-h^{\alpha\sigma,0}h_{\alpha\sigma,0}+2h^{\alpha\sigma}h_{\alpha\sigma,00}),
\end{eqnarray}
where $h=\eta^{\alpha\beta}h_{\alpha\beta}$.  This result mentions
that the integral
$\int_{V}\sqrt{-g}\partial_{0}({}_{M}\tau_{0}{}^{0})d^{3}x$
contains second time derivatives which implies zero tidal heating
value base on the approximation limit (i.e., see
(\ref{23cSep2015})).

\subsection{S pseudotensor}
Here we verify that S does not satisfactory be an acceptable energy-momentum pseudotensor.  Though $S$ obtains the desired tidal
heating value, it fails the expected inside matter result. We
define the superpotential for S as follows
\begin{eqnarray}
_{S}U_{\alpha}{}^{[\mu\nu]}:=\sqrt{-g}(\Gamma^{\beta\tau}{}_{\beta}-\Gamma^{\tau\beta}{}_{\beta})
\delta^{\rho}_{\alpha}\delta^{\mu\nu}_{\rho\tau}.
\end{eqnarray}
In fact, one can rewrite
$_{S}U_{\alpha}{}^{[\mu\nu]}={}_{F}U_{\alpha}{}^{[\mu\nu]}-{}_{M}U_{\alpha}{}^{[\mu\nu]}$
and the pseudotensor S can be represented as
$_{S}t_{\alpha}{}^{\mu}={}_{E}t_{\alpha}{}^{\mu}-{}_{M}t_{\alpha}{}^{\mu}$.
Comparing the results from the previous two subsections, the value for inside matter 
is $2G^{\mu}{}_{\alpha}-R^{\mu}{}_{\alpha}$ in Riemann normal coordinates.
Moreover, the energy
density and tidal heating of S pseudotensor are the same as the
Einstein pseudotensor.

\section{Conclusion}
Thorne verified that tidal heating is independent of how one localizes the gravitational energy and the value is unambiguous.  
Purdue and Favata used different pseudotensors to demonstrate the tidal heating and all of them give the expected result.  However, after re-examined the M$\o$ller pseudotensor, we find that it gives vanishing result which means tidal heating is pseudotensor dependent.
We emphasize that a suitable gravitational energy-momentum
pseudotensor requires fulfill the Freud typed superpoential and this
requirement ensures the expected tidal heating.  Conversely, the
pseudotensor may lose the significance if only succeeded achieving the
disired tidal heating but excluded inside matter value (e.g., 
pseudotensor S).  Therefore collecting the qualified pseudotensor is essential and indeed the tidal heating  becomes pseudotensor independent.



\begin{thebibliography}{3}

\bibitem{Peale}
Peale S J, Cassen P and Reynolds R T 1979 {\it{}Science}
{\bf{}203} 892


\bibitem{Thorne}
Thorne K S 1998 {\it{}Phys. Rev. D} {\bf{}58} 124031

\bibitem{Purdue}
Purdue P 1999 {\it{}Phys. Rev. D} {\bf{}60} 104054


\bibitem{LL}
Landau L D and Lifshitz E M 1975 {\it{}The classical theory of
fields} (Oxford: Pergamon)


\bibitem{Favata}
Favata M 2001 {\it{}Phys. Rev. D} {\bf{}63} 064013



\bibitem{Freud}
Freud Ph 1939 {\it{}Ann. Math.} {\bf{}40} 417



\bibitem{BT}
Bergmann P G and Thomson R 1953 {\it{}Phys. Rev.} {\bf{}89} 400



\bibitem{Moller}
M{\o}ller C 1958 {\it{}Ann. Phys.} NY{\bf{}4} 347


\bibitem{CQGSoNesterChen2009}
So L L, Nester J M and Chen H 2009 {\it{}Class. Quantum Grav.}
{\bf{}26} 085004


\bibitem{SoarXiv}
So L L arXiv:1505.04589


\bibitem{MTW}
Misner C W, Thorne K S and Wheeler J A 1973 {\em{}Gravitation}
(San Francisco, CA: Freeman)



\bibitem{ADM}
Arnowitt R, Deser S and Misner C W 1961 {\it{}Phys. Rev.}
{\bf{122}} 997


\bibitem{CQGSo2009}
So L L 2009 {\it{}Class. Quantum Grav.} {\bf{}26} 185004







\end{thebibliography}
\end{document}